\documentstyle[amscd,amssymb,verbatim,12pt]{amsart}
\theoremstyle{plain}

\theoremstyle{definition}

\numberwithin{equation}{section}

\def\dspace{\baselineskip=0.3 in}
\begin{document}
\dspace
\title[Gravitational Origin of ...]{Gravitational Origin of Phantom Dark
  Energy and late cosmic acceleration }

\author[S.K.Srivastava]%
        {    }

\maketitle

\centerline{\bf S.K.Srivastava }

\centerline{ Department of Mathematics, North Eastern Hill University,}

\centerline{ NEHU Campus,Shillong - 793022 ( INDIA ) }

\centerline{e-mail:srivastava@@nehu.ac.in ; sushil@@iucaa.ernet.in }

\smallskip

\centerline{\bf Abstract}

\smallskip

Here, dark energy is obtained using dual roles of the Ricci scalar
(as a physical field as well as geometry).  Dark
energy density, obtained  in this model, mimics phantom and the derived
Friedmann equation contains a term $\rho_{\rm de}^2/{2\lambda}$ with $\rho_{\rm
de}$ being the dark energy density and $\lambda$, called as cosmic tension. It
is like brane-gravity inspired Friedmann equation, which arises here without
using the brane-gravity theory. It is found that acceleration
is a {\em transient} phenomenon for $\lambda < 0,$ but for $\lambda > 0$ accelerated
expansion is found to encounter the big-rip problem. It is shown that this
problem can be avoided if the dark energy behaves as barotropic fluid and
generalized Chaplygin gas simultaneously.  Moreover, time for transition
(from deceleration to acceleration of the universe) is derived as a function of
equation of state parameter ${\rm w}_{\rm de} = p_{\rm de}/\rho_{\rm de}$ with
$p_{\rm de}$ being the pressure for dark energy fluid.

 

\newpage

\noindent {\bf 1. Introduction} 

Astronomical observations, during the last few years, provide compelling
evidences in favor of accelerating universe at present, which is caused by
dominance  of dark energy (DE) \cite{sp,ar1, ar2,ar3, ar4,na,dn,rs}. Observation
of 16 Type Ia supernovae (SNe Ia) by $Hubble$ $Space$ $Telescope$ further modifies
these results and shows evidence for cosmic deceleration preceding 
acceleration in the late universe  \cite{ag}. So, DE has a significant role in cosmic dynamics of
the current universe. The simplest candidate for DE is supposed to be the
cosmological constant $\Lambda$ which is very high in the early universe, but
there is no mechanism to bring it to present value without
fine-tuning. Alternatively, to explain its decay from a very high value, in
the early universe, to its present extremely small value, many models
\cite{jm,vs} were suggested in which $\Lambda$ is envisaged as a slowly
varying function of cosmic time. Apart from dynamical $\Lambda$ as DE, other
models are fluid-dynamics models (where barotropic fluid is its source with or
without dissipative pressure) , Chaplygin gas(GC) and generalized Chaplygin gas
(GCG)models \cite{rj, ob,mc1, mc2, mc3, mc4,  mc5,sks}. In
field-theoretic models, the most natural ones are models, where DE is caused
by scalars. These are quintessence models \cite{brp1, brp2, brp3, brp4, brp5,   brp6, brp7, brp8, brp9, brp10, brp11, brp12, brp13}, k-essence models
\cite{ca1, ca2}, tachyon models \cite{as1, as2, as3, as4, as5, as6, as7, as8,
  as9,as10} and phantom models \cite{sks, rr1,  rr2,  rr3,  rr4,  rr5, rr6,  rr7,  rr8,  rr9,  rr10,  rr11, rr12}. In
these models, non-gravitational lagrangian density for exotic matter giving DE
is added to Einstein-Hilbert term in the action. Recently, in a different
approach, non-gravitational term is replaced by gravitational term being
non-linear in Ricci scalar $R$, which stems modified gravity
\cite{sc1,sc2,sm,add1,add2, add3,add4,sno1,sno2,sns1,sns2,de1,de2,de3,smc1,smc2,scv1,scv2,scv3,snsd,ms1,ms2,ms3,ms4,ka,om,sn06}. Thus, in these
theories, non-linear term of $R$ is introduced as  DE lagrangian.

In all these models, a non-gravitational or gravitational term for DE is added
in the theory {\it a priori}. In this sense, these models are
phenomenological. Here also, non-linear term of $R$ is added to
Einstein-Hilbert term, but contrary to
\cite{sc1,sc2,sm,add1,add2, add3,add4,sno1,sno2,sns1,sns2,de1,de2,de3,smc1,smc2,scv1,scv2,scv3,snsd,ms1,ms2,ms3,ms4,ka,om,sn06}, in this
model,non-linear term of $R$ is {\em not} used as lagrangian for DE. Rather,
DE density emerges spontaneously, which makes it different from earlier models
of gravitational DE \cite{sc1,sc2,sm,add1,add2, add3,add4,sno1,sno2,sns1,sns2,de1,de2,de3,smc1,smc2,scv1,scv2,scv3,snsd,ms1,ms2,ms3,ms4,ka,om,sn06}. In
the present model , DE emerges as combined effect of linear as well as
non-linear term of $ R$ manifesting dual role of the Ricci scalar as a
physical field as well as geometry. Thus, here, DE density is derived from the
modified gravity. So, this approach of getting gravitational DE is different
from the approach of  \cite{sc1, sc2,sm,add1,add2, add3,add4,sno1,sno2,sns1,sns2,de1,de2,de3,smc1,smc2,scv1,scv2,scv3,snsd,ms1,ms2,ms3,ms4,ka,om,sn06},
where energy-momentum tensor of DE is obtained from non-linear term of
curvature taken as DE lagrangian. Moreover,in what follows, it is found that
dark matter also emerges due to non-linear term of curvature scalar.

Natural units ($\hbar = c = 1$) are used here with GeV as the fundamental
unit, where $\hbar$ and $c$ have their usual meaning. In this unit, it is
found that $1 {\rm GeV}^{-1} = 6.58 \times 10^{-25} {\rm sec}$.

\bigskip

\noindent {\bf 2. Gravitational action, modified Friedmann equation and cosmic cconsequences}

The action
for higher-derivative gravity is taken as 
$$ S_g = \int {d^4x} \sqrt{- g} \Big[ \frac{R}{16 \pi G} - \alpha R^{(2 - r)}
\Big], \eqno(1)$$
where $R$ is the Ricci scalar curvature, $G = M_P^{-2} ( M_P = 10^{19}$ GeV is
the Planck mass). Moreover,  $\alpha $ is a coupling constant having dimension
(mass)$^{ 2 r}$. In what follows, it is shown that $r = 1 + {\rm w}_{\rm de},$
where ${\rm w}_{\rm de} = p_{\rm de}/\rho_{\rm de} < - 1/3$ with $p_{\rm
  de}(\rho_{\rm de})$ being pressure(energy density) for DE fluid. So, $(2 -
r) > 0$ showing that instability problem does not arise \cite{sc1, sc2,sm}.

The action (1) yields gravitational field equations
$$\frac{1}{16 \pi G} ( R_{\mu\nu} - \frac{1}{2} g_{\mu\nu} R ) - \alpha [ (2 -
r) \{
\triangledown_{\mu} \triangledown_{\nu}R^{(1 - r)} - g_{\mu\nu} {\Box} R^{(1 - r)} + R^{(1 - r)}
R_{\mu\nu} \}$$
$$- \frac{1}{2} g_{\mu\nu} R^{(2 - r)}] = 0 \eqno(2)$$
using the condition $\delta S_g/{\delta g^{\mu\nu}} = 0.$ Here, $\triangledown_{\mu}$
denotes covariant derivative and the operator $\Box$ is given as
$${\Box} = \frac{1}{\sqrt{-g}} \frac{\partial}{\partial x^{\mu}}
\Big(\sqrt{-g} g^{\mu\nu} \frac{\partial}{\partial x^{\nu}} \Big) \eqno(3)$$
with $\mu, \nu = 0,1,2,3$ and $g_{\mu\nu}$ as metric tensor components.

Taking trace of eqs.(2) and doing some manipulations, it is obtained that
$${\Box}R  - \frac{r}{R}\triangledown^{\nu}R \triangledown_{\nu}R = \frac{1}{3
  (2 - r)(1 - r)} \Big[\frac{R^{(1 + r)}}{16 \pi G \alpha} - r R^2 \Big]\eqno(4)$$
with $ \alpha \ne 0$ to avoid the ghost problem. 

Experimental evidences support spatially homogeneous flat model of the
universe \cite{ad1, ad2, ad3, ad4, ad5}. So, the line-element, giving geometry of the universe, is
taken as
$$ dS^2 = dt^2 - a^2(t)[dx^2 + dy^2 + dz^2] \eqno(5)$$
with $a(t)$ as the scale factor.

In the homogeneous space-time, given by eq.(5), eq.(4) is obtained as
$$ {\ddot R} + 3 \frac{\dot a}{a}{\dot R} - \frac{r {\dot R}^2}{R} = \frac{1}{3
  (2 - r)(1 - r)} \Big[\frac{R^{(1 + r)}}{16 \pi G \alpha} - r R^2 \Big]\eqno(6)$$

In most of the situations, for example, radiation model, matter-dominated
model, and accelerated models, we have $a(t)$ as a power-law solution yielding
$R$ as the power-law function of $a(t)$ . So, it is reasonable to take $R$
as
$$ R = \frac{A}{a^n}  \eqno(7)$$
with $n > 0$ being a real number and $A$ being a constant with mass
dimension 2.

$R$, given by eq.(7), satisfies eq.(6), if
$$ \frac{\ddot a}{a} + (2 + rn - n) \Big(\frac{\dot a}{a} \Big)^2 = \frac{1}{3
  (2 - r)(1 - r)} \Big[- \frac{a^{-nr}}{8 \pi G n \alpha} + \frac{ r A}{n} a^{-n}\Big] \eqno(8)$$

Eq.(8) integrates to
$$ \Big(\frac{\dot a}{a} \Big)^2 = \frac{C}{a^{2[3 - n (1 - r)]}} +  \frac{1}{3
  (2 - r)(1 - r)} \Big[- \frac{a^{-nr}}{8 \pi G n \alpha ( - nr + 2[3 - n (1 -
  r)]) A^r} $$
$$+ \frac{ r A}{n(- n + 2[3 - n(1 - r)])} a^{-n}\Big] \eqno(9)$$
with $C$ being an integration constant. Setting $n = 3$
eq.(9) yields the modified Friedmann
  equation (MFE)
$$\Big(\frac{\dot a}{a} \Big)^2 = \frac{8\pi G}{3} \Big( \rho_{\rm   de}\Big[1 +
\frac{\rho_{\rm   de}}{2 \lambda}\Big] + \rho_m \Big)  \eqno(10)$$
with
$$\frac{8\pi G}{3}\rho_{\rm   de} = - \frac{  A^r}{216 \pi G \alpha r(2 - r)(1
  - r)} a^{-nr},  \eqno(11)$$

$$\frac{8\pi G}{3} \rho_{\rm m} = \frac{r A}{27(2 - r)(1
  - r) ( -1 + 2r) a^3} \eqno(12)$$
and

$$ \lambda = \frac{A^r}{\Big[ 48 \pi G C \{ 72 \pi G \alpha r (2 - r)(1 -
  r)\}^2 \Big]}. \eqno(13)$$

$\rho_{\rm m}$, given by eq.(12), has the form of matter (pressureless ideal
fluid) density emerging from the gravitational sector. So, it is termed as
density for cold dark matter. Moreover $\rho_{\rm   de}$, given by eq.(11), is also inspired by
modified gravity, so it is called DE density. 

Conservation equation for DE 
$$ {\dot \rho_{\rm de}} + 3 \frac{\dot a}{a} \rho_{\rm de}(1 + {\rm w}_{\rm
    de}) = 0,   \eqno(14a)$$
where equation of state (EOS) parameter 
$${\rm w}_{\rm  de} = p_{\rm  de}/\rho_{\rm de}  \eqno(14b)$$ with $p_{\rm
  de}$ and $\rho_{\rm de}$ being isotropic pressure and density for DE
    respectively. 

Eq.(14a) yields
$$ r  =  (1 + {\rm w}_{\rm  de}) . \eqno(15)$$

Connecting eqs.(12) and (15), it is obtained that $\rho_{\rm m} > 0,$ if
$$ - \frac{(1 + {\rm w}_{\rm de} )}{{\rm w}_{\rm de}(1 - {\rm w}_{\rm de} )(1
  + 2{\rm w}_{\rm de} ) } > 0  \eqno(16)$$
The inequality (16) is possible for
$$ (1 + {\rm w}_{\rm de} ) < 0 \eqno(17a)$$
as DE characterized by ${\rm w}_{\rm de} < - 1/3$. It means that DE, obtained
here, mimics phantom. Thus, eq.(14) and (16) yield
$$ (1 - r) = - {\rm w}_{\rm  de} > 0 .  \eqno(17b)$$

 WMAP data give the present value of $\rho_{\rm de}$ to be $\rho^0_{\rm (de)} = 0.73
 \rho_{\rm cr.}$ and $\rho_{\rm m}$ to be $\rho^0_{\rm m} = 0.23
 \rho_{\rm cr.},$ where $\rho_{\rm cr.} = 3 H_0^2/ 8 \pi G$ with $H_0 =  100 h
 {\rm km/sec Mpc} = 2.33 \times 10^{-42} {\rm GeV}$ and $h = 0.68$ (having
 maximum likelihood) \cite{abl,am03}. Using these values, constants $\alpha$
 and  $A$ in eqs.(11) and (12) are evaluated as
$$ 216 \pi G \alpha (1 - {\rm w}_{\rm de} ) |- {\rm w}_{\rm de} (1 + {\rm
  w}_{\rm de} )| = A^r H_0^{-2} a_0^{3 |1 + {\rm w}_{\rm de}|}  \eqno(18)$$
and
$$ \frac{|1 + {\rm w}_{\rm de}| A}{27(1 - {\rm w}_{\rm de} ) |- {\rm w}_{\rm
  de} (1 + {\rm   w}_{\rm de} )| }  = 0.23 H_0^2  a_0^{3} , \eqno(19)$$
where $a_0 = a(t_0)$ with $t_0 = 13.7 {\rm Gyr} = 6.6 \times 10^{41} {\rm
  GeV}^{-1}$ being the present age of the universe. Here onwards, $a_0$ is
noemalized as
$$ a_0 = 1.  \eqno(20)$$

Incorporating eqs. (15)-(20) in eqs.(10)-(13), it is obtained that
$$\Big(\frac{\dot a}{a} \Big)^2 = \frac{8 \pi G}{3}\Big[\rho_{\rm   de} \Big\{1 +
\frac{\rho_{\rm   de}}{2 \lambda}\Big \} + \rho_{\rm   m} \Big]  \eqno(21a)$$
with
$$\rho_{\rm   de} = 0.73 \rho_{\rm cr} a^{ 3 |1 + {\rm w}_{\rm  de}|},
 \eqno(21b)$$

$$\rho_{\rm   m} = 0.23 \rho_{\rm cr} a^{- 3},
 \eqno(21c)$$

and

$$ \lambda =  \Big[\frac{16}{3} \pi G C\Big]^{-1} H_0^4   \eqno(21d)$$

It is interesting to
see that MFE (10)  contains a term $\rho_{\rm   de}^2/{2
  \lambda}$, which is non-linear in $\rho_{\rm   de}$. As a reference, it is
right to mention that brane-gravity inspired Friedmann equation also has a
non-linear term like $\rho^2/2\lambda_b$ with $\rho$ being the energy density
and $\lambda_b$ being the brane-tension \cite{rm}. Here, MFE (10) arises from
the modified gravity without using any prescription for the brane-gravity and
$\lambda$ is termed as {\em cosmic tension}.
 
In eq.(10), energy density is a physical concept, which is caused by the
scalar curvature $R$. It shows role of $R$ as a physical field also. So, it is
reasonable to remark that $R$ has dual role here (i) as a physical field and
(ii) as a geometrical field \cite{ks1,ks2, ks3, ks4, ks5, ks6, ks7, ks8, ks9,
  aas1,aas2, aas3,  skp1,skp2, skp3, skp4, skp5, skp6, skp7, skp8, skp9,
  skp10, skp11, skp12}. 

$\lambda$, given by (21d), can be positive as well as negative. So (21a) is
re-written as

$$\Big(\frac{\dot a}{a} \Big)^2 = \frac{8 \pi G}{3}\Big[\rho_{\rm   de} \Big\{1 \pm
\frac{\rho_{\rm   de}}{2 |\lambda|}\Big \} + \rho_{\rm   m} \Big]
. \eqno(22)$$
with $(+)$ sign for $\lambda > 0$ and $(-)$ sign for $\lambda < 0$.
In case, $\lambda < 0,$ effect of phantom DE vanishes when $\rho_{\rm   de}$
grows to  $2 |\lambda|$ in future and accelerated expansion will come to an end.

Now, $\rho_{\rm m} > \rho_{\rm de}$ when red-shift $z$ is given as
$$ z > \Big(\frac{73}{23} \Big)^{1/3|{\rm w}_{\rm de}|} - 1. \eqno(23a)$$

 $\rho_{\rm m} < \rho_{\rm de}$  for
$$ z < \Big(\frac{73}{23} \Big)^{1/3|{\rm w}_{\rm de}|} - 1. \eqno(23b)$$

Inequalities (23a) and (23b) show that transition from $\rho_m > \rho_{\rm de}$
to $\rho_{\rm m} < \rho_{\rm de}$ takes place at
$$z_* = \Big(\frac{73}{23} \Big)^{1/3|{\rm w}_{\rm de}|} - 1 =
\Big(\frac{a_0}{a_*} \Big) - 1. \eqno(24)$$ 

Observations yield $-1.22 < {\rm w}_{\rm de} < - 1$ \cite{ag} for phantom DE. Using these values of ${\rm w}_{\rm de}$ , (24) yields
$$ 0.37 < z_* \lesssim 0.46  ,  \eqno(25)$$
which is supported by the range $ 0.33 \lesssim  z_* \lesssim 0.59$ given by
16 Type Supernova observations \cite{ag}.

When $z > z_*, \rho_{\rm m} > \rho_{\rm de}$ and $ \rho_{\rm de}/2|\lambda| <
1,$ so (22) reduces to
$$\Big(\frac{\dot a}{a} \Big)^2 = 0.23 H_0^2 \Big(\frac{a_0}{a}
\Big)^3,  \eqno(26)$$ 
using (21c) for $\rho_{\rm m}.$

(26) integrates to 
$$ a(t) = a_d \Big[ 1 + 0.72 H_0 \Big(\frac{1}{a_d} \Big)^{3/2} (t - t_d)
\Big]^{2/3}  \eqno(27)$$
with $a_d = a(t_d)$ being a constant.$ a(t) $, given by (27), shows deceleration.

Further, for  $z < z_*, \rho_m < \rho_{\rm de}$, so (22) reduces to
$$\Big(\frac{\dot a}{a} \Big)^2 = 0.73 H_0^2 a^{3 |1 +
  {\rm w}_{\rm de}|}\Big [1 \pm
\frac{\rho_{\rm   de}}{2 |\lambda|}\Big ] ,  \eqno(28)$$ 

\bigskip

\noindent {\bf Case 1. When $\lambda < 0$}

In this case, (28) looks like
$$\Big(\frac{\dot a}{a} \Big)^2 = 0.73 H_0^2 a^{3 |1 +
  {\rm w}_{\rm de}|}\Big [1 -
\frac{\rho_{\rm   de}}{2 |\lambda|}\Big ] ,  \eqno(29)$$ 
which integrates to

$$ a(t) = \Big[\frac{0.73 \rho_{\rm cr}}{2 |\lambda|} + \Big\{a_*^{-3 |1 +
  {\rm w}_{\rm de}|/2} \sqrt{1 - \frac{0.73 \rho_{\rm cr}a_*^{3 |1 +
  {\rm w}_{\rm de}|}}{2 |\lambda|}}$$
$$ - \frac{3 |1 +   {\rm w}_{\rm de}|}{2} H_0 \sqrt{0.73} (t - t_*) \Big\}^2
\Big]^{- 1/3 |1 +   {\rm w}_{\rm de}|} \eqno(30)$$ 
giving accelerated expansion, which will continue tiil the time $t_{\rm e}$
such that 
$$ \rho(t_{\rm e} ) = 2 |\lambda|.$$ 
implying
$$0.73 \rho_{\rm cr} a_{\rm e}^{3 |1 +   {\rm w}_{\rm de}|} = 2 |\lambda|.\eqno(31)$$

Connecting (30) and (31), it is obtained that
$$ t_{\rm e} - t_* = \frac{2}{3 |1 +   {\rm w}_{\rm de}|H_0} \sqrt{(1 +
  z_*)^{3 |1 +   {\rm w}_{\rm de}|} - \frac{0.73 \rho_{\rm cr}}{2 |\lambda|}} \eqno(32)$$
giving period of acceleration.

The  time $t_*$ for transition from deceleration to acceleration is obtained
as

$$ t_* = t_0 - \frac{2}{3|1 +   {\rm w}_{\rm de}| H_0
  \sqrt{0.73}}\Big[\sqrt{(1 + z_*)^{3|1 +   {\rm w}_{\rm de}|} - \frac{0.73
  \rho_{\rm cr}}{2 |\lambda|}} - \sqrt{ - \frac{0.73 \rho_{\rm cr}}{2
  |\lambda|}}\Big] \eqno(33)$$

For $t > t_{\rm e}, \rho_{\rm de} > 2 |\lambda| = \rho(t_{\rm e} )$ and (22) takes the form

$$\Big(\frac{\dot a}{a} \Big)^2 = \frac{8 \pi G}{3}\Big[- \rho_{\rm   de}
\Big\{ \frac{\rho_{\rm   de}}{2 |\lambda|} - 1\Big \} + \rho_{\rm   m} \Big]
. \eqno(34)$$

It is interesting to see that (34) yields ${\dot a} = 0, $ at some future time
$t = t_{\rm m}$,when we have
$$ \frac{\rho_{\rm m}(t_{\rm m})}{\rho_{\rm de}(t_{\rm m})} = \frac{\rho_{\rm de}(t_{\rm m})}{2 |\lambda|} - 1
= \frac{\rho_{\rm de}(t_{\rm m})}{\rho_{\rm de}(t_{\rm e})} - 1 \eqno(35)$$
and $a(t)$ grows to its maximum $a_{\rm m}$ satisfying the condition

$$a_{\rm e}^{3 (1 + {\rm w}_{\rm de})} = a_{\rm m}^{3 (1 + {\rm w}_{\rm de})}
+ \frac{23}{73} a_{\rm m}^{3 (1 + 2{\rm w}_{\rm de})} \eqno(36)$$
for $(1 + {\rm w}_{\rm de}) < 0.$

During the period $(t_{\rm m} - t_{\rm e}),$ (34) is effectively like (26)
yielding decelerated expansion, driven by matter, as 
$$ a(t) = a_{\rm e} \Big[ 1 + \frac{3 H_0}{2 a_{\rm e}^{3/2}} \sqrt{0.23} (t -
t_{\rm e}) \Big]^{2/3} . \eqno(37)$$ 

\bigskip

\noindent {\bf Case 2. When $\lambda > 0$}

In this case, (28) looks like
$$\Big(\frac{\dot a}{a} \Big)^2 = 0.73 H_0^2 a^{3 |1 +
  {\rm w}_{\rm de}|}\Big [1 +
\frac{\rho_{\rm   de}}{2 |\lambda|}\Big ] ,  \eqno(38)$$ 
which integrates to

$$ a(t) = \Big[- \frac{0.73 \rho_{\rm cr}}{2 |\lambda|} + \Big\{a_*^{-3 |1 +
  {\rm w}_{\rm de}|/2} \sqrt{1 + \frac{0.73 \rho_{\rm cr}a_*^{3 |1 +
  {\rm w}_{\rm de}|}}{2 |\lambda|}}$$
$$ - \frac{3 |1 +   {\rm w}_{\rm de}|}{2} H_0 \sqrt{0.73} (t - t_*) \Big\}^2
\Big]^{- 1/3 |1 +   {\rm w}_{\rm de}|} \eqno(39)$$ 
showing $a \to \infty,$ when 
$$t \to t_{\rm s} = t_* + [1.095 |1 + {\rm w}_{\rm de}| H_0 ]^{-1} \times $$
$$\Big[\sqrt{(1 + z_*)^{3 |1 + {\rm w}_{\rm de}|} + \frac{0.73 \rho_{\rm cr}}{2
      |\lambda|}} \pm \sqrt{\frac{0.73 \rho_{\rm cr}}{2
      |\lambda|}} \Big] .  \eqno(40)$$
Moreover, as $t \to t_{\rm s}, \rho_{\rm de} \to \infty$ and $p_{\rm de} \to -
\infty$ showing the big-rip problem.

\bigskip

\noindent {\bf 3. Avoidance of big-rip singularity}

In what follows, it is shown that the ``big-rip singularity'',
obtained above for positive cosmic tension $\lambda$ can be avoided if DE
(obtained here) as a barotropic fluid obeying (14b) as well as generalized
Chaplygin gas (GCG) with EOS given as \cite{rj,ob,mc1, mc1,mc3, mc4, mc5,sks}
$$ p_{\rm de} = - \frac{M^{1 + \beta}}{\rho_{\rm de}^{\beta}}, \eqno(41)$$  
where $0 < \beta < 1$ for GCG , $\beta = 1$ for CG and $M$ is a constant.

Due to negative pressure and being only fluid with supersymmetric
generalization , GC and GCG have been strong candidates for DE. But
experimental results support GCG \cite{ob,mc1, mc1,mc3, mc4, mc5}. So, GCG model is
preferred here. Avoidance of ``big-rip singularity'', using double behaviour of DE
fluid as being done here, was explored in \cite{sks} for non-gravitational
phantom DE.

(14b) and (41) yield
$$ {\rm w}_{\rm de} = - \Big(\frac{M}{\rho_{\rm de}} \Big)^{1 +
  \beta}. \eqno(42)$$ 
This equation shows dependence of $ {\rm w}_{\rm de}$ on $\rho_{\rm de}$
  varying with $a(t)$. Now, using eq.(42) for $\rho_{\rm de}$ with variable $
  {\rm w}_{\rm de}$, conservation equation (14a) looks like
$$ {\dot {\rm w}_{\rm de}} = 3 \frac{\dot a}{a } {\rm w}_{\rm de} (1 + {\rm
  w}_{\rm de} ), \eqno(43)$$ 
which integrates to
$$ \frac{{\rm w}_{\rm de}}{1 + {\rm w}_{\rm de}} = \frac{{\rm w}^0_{\rm de}}{1
  + {\rm w}^0_{\rm de}} a^{3 (1 + \beta)} \eqno(44)$$
with ${\rm w}^0_{\rm de} = {\rm w}_{\rm de}(a_0)$.

Connecting (42) and (44)

$$ \rho_{\rm de} = M \Big[ 1 - \frac{(1 + {\rm w}^0_{\rm de})}{{\rm w}^0_{\rm
    de}} a^{- 3 (1 + \beta)} \Big]^{1/(1 + \beta)}, \eqno(45)$$
where
$$ M = 0.73 \rho_{\rm cr} |{\rm w}^0_{\rm de}|^{1/(1 + \beta)} \eqno(46)$$
being obtained from (42).

Now, for $\lambda > 0$ and $\rho_{\rm m} < \rho_{\rm de},$ (22) takes the form
$$\Big(\frac{\dot a}{a} \Big)^2 = \frac{8 \pi G}{3}\rho_{\rm   de} \Big[1 +
\frac{\rho_{\rm   de}}{2 |\lambda|}\Big]
. \eqno(47)$$

Connecting (45) and (47) and making approximations, it is obtained that
$$\frac{\dot a}{a} \simeq \sqrt{\frac{8 \pi G}{3} M \Big(1 + \frac{M}{2
    |\lambda|} \Big)} \Big[1 - \frac{(1 + {\rm w}^0_{\rm de})}{2 {\rm
    w}^0_{\rm de} } a^{ - 3(1 + \beta)} \Big], \eqno(48)$$
which integrates to
$$ a(t) = \Big[\frac{(1 + {\rm w}^0_{\rm de})}{2 {\rm
    w}^0_{\rm de} } + \Big\{(1 + z_*)^{-3(1 + \beta)} - \frac{(1 + {\rm
    w}^0_{\rm de})}{2 {\rm  w}^0_{\rm de} } \Big\} \times$$
$$ e^{3(1 + \beta)\sqrt{\frac{8 \pi G}{3} M \Big(1 + \frac{M}{2
    |\lambda|} \Big)}} (t - t_*) \Big]^{1/3(1 + \beta)}, \eqno(49)$$
which is valid for
$$(1 + z_*)^{-3(1 + \beta)} > \frac{(1 + {\rm
    w}^0_{\rm de})}{2 {\rm  w}^0_{\rm de} } . \eqno(50)$$
(49) is free from any finite time singularity listed in \cite{sn05}. Using
(49) in (45) and (42), it is obtained that $\rho_{\rm de} \to M$ and ${\rm
 w}_{\rm de} \to - 1$.

\bigskip

\noindent {\bf 4. Summary}

In summary, it is interesting to see that phantom like dark energy emerges
  spontaneously from the gravitational sector and the Friedmann
  equation(giving cosmic dyanamics) contains a term $\rho_{\rm de}^2/2
  \lambda$ with $\rho_{\rm  de}$ being the dark energy density and $\lambda$ being the cosmic tension
  . It is found that time of transition from deceleration to acceleration
  depends on ${\rm w}_{\rm de}$. Here, two cases arise (i) when cosmic tension
  $\lambda < 0$ and (ii) when cosmic tension $\lambda > 0$. In case of
  negative cosmic tension, late cosmic acceleration begins from the transition
  time $t_*$ and stops at a finite time $t_s$ in future. It shows that the
  negative cosmic tension puts a {\em brake} on accelerated expansion at $t =
  t_s$. Later on, universe decelerates driven by matter showing another
  transition from acceleration to deceleration. Decelerated expansion contiues
  till $a(t)$ acquires its maximum at $t= t_m > t_s$ and the ratio $
  {\rho_{\rm m}(t_{\rm m})}/{\rho_{\rm de}(t_{\rm m})}$ is given by (35).

When cosmic tension is positive, universe accelerates from the time of
transition and ends up into `big-rip'singularity in future. It is shown that
this problem can be avoided if dark energy behaves as a barotropic fluid and
GCG simultaneously. This type of approach was adapted  in the
case of non-gravitational phantom led cosmological model in \cite{sks} using
classical mechanics. It is in contrast to approach in \cite{ee1, ee2, ee3}, where it is
suggested that ``sudden future singularity'' may be avoided using quantum
gravity effects.

\end{document}